\documentclass[12pt]{article}
\usepackage{graphicx, amsmath, amssymb, cite, setspace, color}

\usepackage[top=1 in, bottom=1 in, left=1 in, right=1 in]{geometry}

\newcommand{\captionfonts}{\footnotesize} 
\makeatletter 
\long\def\@makecaption#1#2{%
  \vskip\abovecaptionskip
  \sbox\@tempboxa{{\captionfonts #1: #2}}%
  \ifdim \wd\@tempboxa >\hsize
    {\captionfonts #1: #2\par}
  \else
    \hbox to\hsize{\hfil\box\@tempboxa\hfil}%
  \fi
  \vskip\belowcaptionskip}
\makeatother 

\def\fnote#1#2{\begingroup\def\thefootnote{#1}\footnote{#2}
     \addtocounter{footnote}{-1}\endgroup}

\begin{document}
\title{Bubbles of Nothing  and the \\ Fastest  Decay in the Landscape}
\author{Adam~R.~Brown$^{1,2}$ and Alex~Dahlen$^{1}$ \vspace{.1 in}\\  
\vspace{-.3 em}  $^1$ \textit{\small{Physics Department, Princeton University, Princeton, NJ 08544, USA}}\\
\vspace{-.3 em}  $^2$ \textit{\small{Princeton Center for Theoretical Science, Princeton, NJ 08544, USA}}}
\date{}

\maketitle
\fnote{}{emails: \tt{adambro@princeton.edu, adahlen@princeton.edu}}
\vspace{-.95cm}

\begin{abstract}
\noindent The rate and manner of vacuum decay are calculated in an explicit flux compactification, including all thick-wall and gravitational effects.  For landscapes built of many units of a single flux, the fastest decay is usually to discharge just one unit.  By contrast, for landscapes built of a single unit each of many different fluxes, the fastest decay is usually to discharge all the flux at once,  which destabilizes the radion and begets a bubble of nothing. By constructing the bubble of nothing as the limit in which ever more flux is removed, we gain new insight into the bubble's appearance.  Finally, we describe a new instanton that mediates simultaneous flux tunneling and decompactification. Our model is the thin-brane approximation to six-dimensional Einstein-Maxwell theory.
\end{abstract}

\section{Introduction}

Theories with extra dimensions typically give rise to large landscapes of vacua stabilized by flux.  This means that most minima have a great many potential decay paths: they may decay by a \emph{small step} (dropping just one or a few units of flux); they may decay by a \emph{giant leap} (dropping many units of flux); they may decay by the giantest leap of all, a \emph{bubble of nothing} (dropping every unit of flux, so that the extra dimensions collapse to zero size and spacetime pinches off); they may \emph{decompactify} (so that the extra dimensions run away to infinite size); or they may simultaneously decompactify \emph{and} drop flux. In this paper we will construct the tunneling solutions that describe these processes and compare their relative rates, in the context of an explicit flux compactification.

In a previous paper \cite{paper1}, we described effects that generically enhance the rate of giant leaps and showed that \emph{in the thin-wall approximation} indeed giant leaps can be the fastest decay route for `monoflux' landscapes---one-dimensional landscapes built of many units of a single flux. Then, in \cite{paper2}, we described additional enhancements specific to `multiflux' landscapes---multi-dimensional landscapes built of a single unit each of many different fluxes. In Section \ref{sec:6dEM}, we review these arguments, and introduce our explicit flux compactification: 6D Einstein-Maxwell theory compactified on a two-sphere.   Transitions occur by the nucleation of a charged black two-brane; specifically we assume there are no fundamental charged branes of lower tension.  In Section \ref{sec:fluxinstantons},  we construct the pertinent instantons, including all thick-wall effects, though still treating the black brane as thin. In Section \ref{sec:Bofn}, we calculate the decay rate as a function of the distance jumped. In monoflux landscapes, we find that small steps beat giant leaps because thick-wall effects overwhelm the enhancements described in \cite{paper1}. However, in multiflux landscapes,  giant leaps beat small steps from almost all starting points, to such an extent that the fastest decay is to dump all the flux at once, jump over the entire landscape, and nucleate a bubble of nothing.

In Section \ref{sec:BubbleOfNothing}, we examine the bubble of nothing in detail and show how it is approached as the limit of flux tunneling in which all units of flux are discharged. In this limit, not only does the size of the extra dimensions go to zero inside the bubble, but the interior becomes so negatively curved that a 3D slice through the bubble has surface area but no volume.
(Witten's original bubble of nothing was constructed for an \emph{unstabilized} extra dimension \cite{Witten}; ours is for  \emph{stabilized} extra dimensions.) In the Appendix, we describe a new decay that combines flux tunneling and decompactification, and argue that it is always subdominant.   

(We study, in this paper, transitions out of the four-dimensional vacua; others have considered transitions out of the six-dimensional vacuum \cite{ChangeD}. Our enhancements for stacks of branes lie not in the prefactor \cite{Feng:2000if} but in the exponent. Despite the enhancements, the fastest decay is still generically exponentially suppressed, so we don't expect percolation \cite{perc1,perc2}.)

\section{Review of 6D Einstein-Maxwell theory} \label{sec:6dEM}

In this section, we review a simple flux compactification, 6D Einstein-Maxwell theory \cite{FreundRubin}, and flux tunneling between the corresponding vacua \cite{BSV}. The six-dimensional action is
\begin{equation}
S_\text{EM}= \int d^{\,6}\!\,x \, \sqrt{-G}\left(\frac12 \mathcal{R}^{(6)}-\frac14 F_{AB}F^{AB}-\Lambda_6\right), \label{eq:EMaction}
\end{equation}
where $A$ and $B$ run from $0$ to $5$, $\mathcal{R}^{(6)}$ is the Ricci scalar associated with the metric $G_{AB}$, $F_{AB}$ is the Maxwell field strength, $\Lambda_6$ is a positive six-dimensional cosmological constant, and we use units for which $\hbar$, $c$, and the reduced 6D Planck mass are all 1.

Two of the dimensions are compactified on a sphere, which is buttressed against collapse by flux; specifically $\mathfrak{N}$ different two-form fluxes, each with field strength
\begin{equation}
\mathbf{F}_i=\frac{g_i N_i}{4\pi} \sin \theta \,  d\theta \wedge d \phi,
\end{equation}
where $i$ runs from $1$ to $\mathfrak{N}$, $g_i$ is the charge of the $i$th magnetic quantum, $N_i \in \mathbb{N}$ is the number of units of the $i$th magnetic flux, and $\theta$ and $\phi$ span the extra-dimensional two-sphere. We can choose coordinates
\begin{equation}
ds^2=e^{-\psi(x)/M} g_{\mu\nu}dx^\mu dx^\nu + e^{\psi(x)/M} R^2  d\Omega_2^{\, 2} \label{EinsteinFrameCoordinates}
\end{equation}
such that after integrating out the extra dimensions we end up in Einstein frame:
\begin{equation}
S=\int d^4x\sqrt{-g}\left(\frac12 M^{\,2}\mathcal{R}^{(4)}-\frac12\partial_\mu\psi\partial^\mu\psi -V(\psi)\right),
\end{equation}
where the reduced 4D Planck mass is $M=\sqrt{4 \pi} R$, and $\psi$ is the radion. We can shift $\psi$ to set $R=1/\sqrt{2 \Lambda_6}$, then 
\begin{equation}
\label{effpot}
V(\psi) = 4 \pi \left( \frac{1}{2} \frac{F^2}{4 \pi M^2}e^{-3\psi/M}-e^{-2\psi/M}+ \frac{1}{2} e^{-\psi/M} \right),
\end{equation}
and 
\begin{equation}
F^2 = \sum_{i=1}^{\mathfrak{N}} g_i^2 N_i^2. \label{eq:Fsquared}
\end{equation}
The first term in the potential is from flux; the second is from the curvature of the two-sphere; and the third, is from the bulk cosmological constant.
The minimum of this potential is at $\exp [ -  \psi_{\text{min}} / M]  = \frac{2}{3} \frac{4 \pi M^2}{ {F}^2} \left(1 + \sqrt{1 - \frac{3}{4} \frac{ F^2}{4 \pi M^2}  } \right)$,  so 
\begin{equation}
V_{\text{min}} = \frac{4 \pi}{3} \frac{4 \pi M^2}{F^2} \left[ 1 -  \frac{8}{9} \frac{4 \pi M^2}{F^2}  \left(1 + \left[1 - \frac{3}{4} \frac{ F^2 }{4 \pi M^2}  \right]^{\frac{3}{2}} \right) \right]. \label{eq:Vmin}
\end{equation}

Different values of $F^2$ give rise to anti-de Sitter ($F^2 < 4 \pi M^2$), Minkowski ($F^2 = 4 \pi M^2$) or de Sitter ($F^2 > 4 \pi M^2$) four-dimensional vacua, as shown in Fig.~1; as $F^2$ increases, the minimum moves to larger values of $\psi$, and eventually disappears ($F^2 > 16 \pi M^2/3$). 
\begin{figure}[htbp] 
   \centering \label{fig:Vofpsi}
   \includegraphics[width=4.5in]{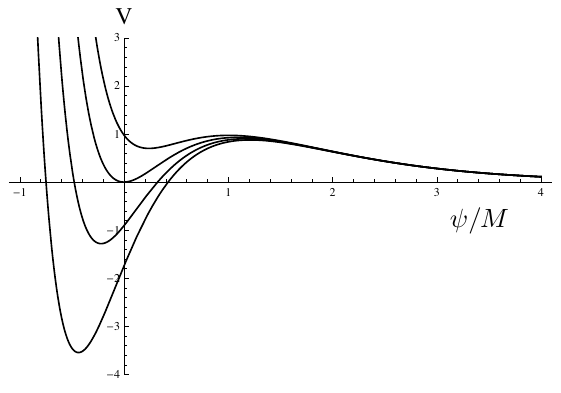}  
   \caption{The effective potential $V(\psi)$ is plotted for different values of $F^2$. Large $F^2$ gives rise to a de Sitter minimum, with $V_\text{min}$ and $\psi_\text{min}$ positive. These minima can decay either by decompactifying (tunneling over the barrier and rolling $\psi \rightarrow \infty$) or by flux tunneling to lower $F^2$. As $F^2$ decreases, the minimum first becomes Minkowski, with $V_\text{min} = \psi_\text{min}=0$, and then AdS, with $V_\text{min}$ and $\psi_\text{min}$ negative; these minima can decay only by flux tunneling. As  $F^2 \rightarrow 0$, the potential destabilizes and $\psi$ can roll towards minus infinity.  This corresponds to a bubble of nothing, as we will see in Sec.~\ref{sec:BubbleOfNothing}.}
\end{figure}e

\subsection{Transitions in 6D Einstein-Maxwell theory}
Transitions between the minima of 6D Einstein-Maxwell theory are effected by the nucleation of a charged two-brane that forms a sphere in the large dimensions and sits at a point in the extra dimensions. 
The brane unravels flux, so that $\psi$ relaxes to a lower minimum of the effective potential on the inside of the sphere. If the brane is charged under $n_i$ units of each flux, then the internal field is reduced $N_i \rightarrow N_i - n_i$.   

The fastest decay uses the lightest brane with a given charge.  Since the theory contains no fundamental branes,  these are extremally charged black two-branes.
The tension of such a brane is
\begin{equation}
T=\frac{2}{\sqrt3} \left( \sum_{i=1}^{\mathfrak{N}} g_i^2 n_i^2 \right)^{\frac{1}{2}} \label{eq:TofgN}.
\end{equation}

In \cite{paper1} we saw two factors that enhance the rate of giant leaps, both caused by the radion.  

The first relates to the brane tension, which, in the thin-brane limit, contributes to the action a term
\begin{equation}
\label{eq:brane}
S_\text{brane} =- T\int_\Sigma \sqrt{-\gamma} d^3\xi= -T e^{-3 \psi(\xi)/2 M} \int_\Sigma \sqrt{-g} d^3\xi, 
\end{equation}
where $\gamma$ is the induced metric.  The brane couples to the radion in such a way that, as the extra dimensions swell, the effective tension of the brane redshifts: this means $\psi$ spikes in the vicinity of branes \cite{Yang}, and the larger the tension, the bigger the spike. This brane-radion coupling induces an attractive force between branes---stacks of branes bind together, lowering their total tension and making them easier to nucleate. 

The second relates to the change in the effective potential $\Delta V_\text{min}$. When flux is discharged, $\Delta V$ at fixed $\psi$ changes linearly with $\Delta F^2$, as in Eq.~\ref{effpot}, but the minimum of the potential also relaxes to a more negative $\psi$, so that $\Delta V_\text{min}$ scales faster than linearly with $\Delta F^2$. Bigger $\Delta V_\text{min}$ means the true vacuum is more preferred and the false vacuum decays faster. 

To illustrate the range of behaviors, we focus on two extreme examples.  {\bf Monoflux landscapes} we define as having only a single type of flux ($\mathfrak{N} =1$) and many units of it ($N \gg 1$), so that $F^2=g^2N^2$.  Nucleating a stack of $n$ branes sends $N\rightarrow N-n$ and the branes are charged under the same flux.  {\bf Multiflux landscapes} we define as having many types of flux ($\mathfrak{N} \gg 1$), but only a single unit of each ($N_i=1$), each with the same charge ($g_i = g$), so that $F^2=g^2\mathfrak N$.  Nucleating a stack of $n$ branes sends $\mathfrak N\rightarrow\mathfrak N-n$ and the branes are now charged under different fluxes.

In \cite{paper2} we saw two additional factors that enhance the rate of giant leaps, this time specific to multiflux landscapes. 

The first relates to Eq.~\ref{eq:TofgN} and the way the tension of a stack of $n$ branes grows with $n$. Monoflux branes all carry the same type of charge, so $T \sim n$. This is because, ignoring the radion, these extremal monoflux branes do not interact---their gravitational attraction is precisely cancelled by their magnetic repulsion, and the tensions add. Multiflux branes are charged under different fluxes, so $T \sim \sqrt{n}$. This is because the branes bind together---there is now no magnetic repulsion to cancel their gravitational attraction.

The second relates to Eq.~\ref{eq:Fsquared} and the way the flux dropped grows with $n$.  For monoflux landscapes, the flux lines are the same and thus repel, so that $\Delta F^2 = g^2 N^2 - g^2(N-n)^2 = g^2 (2n N - n^2)$, which grows slower than linearly with $n$. For multiflux landscapes, the flux lines are different and thus do not interact, so that \mbox{$\Delta F^2 =  g^2 \mathfrak{N} - g^2 (\mathfrak{N}-n) = g^2 n$}, which grows linearly with $n$. 

For given true and false vacuum energies, therefore, the tension of the multiflux brane is larger than that of the monoflux brane:
\begin{eqnarray}
\label{Tmono}
T_\text{mono} &=& \frac2{\sqrt3}\left[(F_\text{false}^2)^{1/2}-(F_\text{true}^2)^{1/2}\right] \\
\label{Tmulti}
T_\text{multi} &=& \frac 2{\sqrt3} \left(F_\text{false}^2-F_\text{true}^2\right)^{1/2}.
\end{eqnarray}
(The tensions agree for $F_\text{true}=0$, the bubble of nothing.)  On account of their larger tension, all multiflux decays are suppressed relative to their monoflux counterparts,  but the least relatively suppressed are the ones with big $n$, the giant leaps.

\section{Flux tunneling instantons} \label{sec:fluxinstantons}
The instantons that describe bubble nucleation can be found by passing to Euclidean signature and imposing O(4) symmetry. The 4D metric can then be written as
\begin{equation}
g_{\mu \nu} dx^{\mu} dx^{\nu} = d \rho^2 + a(\rho)^2 d \Omega_3^{\,2},
\end{equation}
and the action is
\begin{eqnarray}
S_\text{E}&=&2 \pi^2 \int_0^{\bar{\rho}} \! d\rho \left[a^3 \left( \frac12 \psi'^{\,2}+V_{N-n}(\psi)\right)-3M^2a\Bigl(1+(a')^2\Bigl)\right] \nonumber \\[12pt]
 &+& 2\pi^2 a^3  T e^{-3\psi/2 M }{\Big{|}}_{\bar{\rho}} + 2 \pi^2 \int_{\bar{\rho}}^{\,}  d\rho \left[ a^3 \left( \frac12 \psi'^{\,2}+V_N(\psi)\right)- 3M^2a\Bigl(1+(a')^2\Bigl)\right],
\end{eqnarray}
where we have integrated by parts, so that the action only contains single derivatives (and there is no Gibbons-Hawking term).  The action has three pieces, corresponding, respectively, to the bubble's interior, the brane, and the bubble's exterior.  The equations of motion are the Euler-Lagrange equation for $\psi(\rho)$,
\begin{equation}
\psi''+3\frac{a'}{a} \psi'=\frac{dV(\psi)}{d \psi} , \label{eompsi}
\end{equation}
and the gravitational constraint equation for $a(\rho)$,
\begin{equation}
\label{eomgrav}
a'^2=1+\frac{1}{3M^2} a^2 \left(\frac12 \psi'^{\,2}-V(\psi)\right),
\end{equation}  
where $V(\psi)$ is chosen appropriately for inside the brane ($\rho < \bar{\rho}$) and outside the brane ($\rho > \bar{\rho}$). At the brane, there are two boundary conditions, one from varying $\psi(\bar{\rho})$
\begin{equation}
\Delta \psi' \big{|}_{\bar{\rho}} =-\frac32 T e^{-3 \psi/2M} \big{|}_{\bar{\rho}},
\end{equation}
and the other (the Israel junction condition) from varying $a(\bar{\rho})$,
\begin{equation}
\label{bc2}
\Delta a' \big{|}_{\bar{\rho}} = -\frac{a}{2 M^2}T e^{-3 \psi/2M} \big{|}_{\bar{\rho}}.
\end{equation}
This second condition can be better understood by applying Eq.~\eqref{eomgrav} and taking the non-gravitational limit ($M \rightarrow \infty$):
\begin{equation}
\Delta(\frac12 \psi'^{\,2}-V(\psi)) \Big{|}_{\bar{\rho}} = -\frac{3T}{\bar\rho} e^{-3\psi/2M} \Big{|}_{\bar{\rho}} .
\end{equation}
This is a force-balance equation, balancing the pressure differential, on the left, against the surface tension, on the right. For larger $\bar\rho$, the pressure term wins and the bubble can lower its action by expanding; for smaller $\bar{\rho}$, the surface tension term wins and the bubble can lower its action by contracting---this mode is therefore associated with the negative eigenvalue.

At the origin $a(0)=0$ and the instanton must be smooth,
\begin{equation}
\label{regular}
\psi'(0)=0,
\end{equation}
which implies that $a'(0)=1$. (Except for the bubble of nothing, which satisfies its own smoothness condition.) For decay from Minkowski or AdS space, the instanton is infinite, $a(\rho)$ grows without bound, and $\psi(\rho)$ must asymptote to its false vacuum value far from the bubble. For decay from de Sitter space, the instanton is compact, $a(\rho)$ has a second zero at the antipode to $\rho=0$, where, again by smoothness, $\psi' = 0$, though $\psi$ need not sit in its false vacuum value. 

 \begin{figure}[h] 
    \centering
    \includegraphics[width=6.5in]{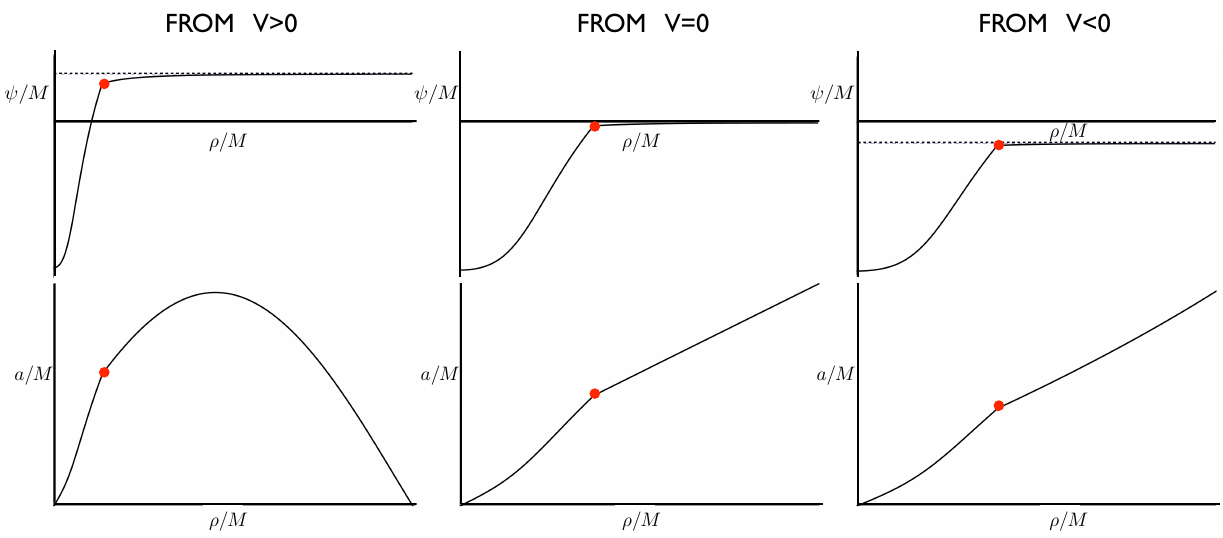} 
    \caption{Sample instanton profiles for flux tunneling in monoflux landscapes. All three transitions end in the same AdS true vacuum, but start from different false vacua, respectively: de Sitter ($ \psi, V>0$), Minkowski ($\psi=V=0$) and AdS ($\psi,V<0$). The radion $\psi(\rho)$ is near its true vacuum value at $\rho = 0$ and from there interpolates smoothly towards its false vacuum value (indicated by the dotted line), except for a derivative discontinuity at the brane (indicated by the red dot). In the de Sitter case, the instanton is compact, so $\psi(\rho)$ never quite reaches its false vacuum value. Deep inside the bubble, $a(\rho)$ is concave up, as is appropriate to AdS; at large $\rho$, $a(\rho)$ behaves like  $\sin \rho$, $\rho$, and $\sinh \rho$ respectively.}
    \label{SampleInstantonMono}
 \end{figure}

 \begin{figure}[h!] 
    \centering
    \includegraphics[width=6.5in]{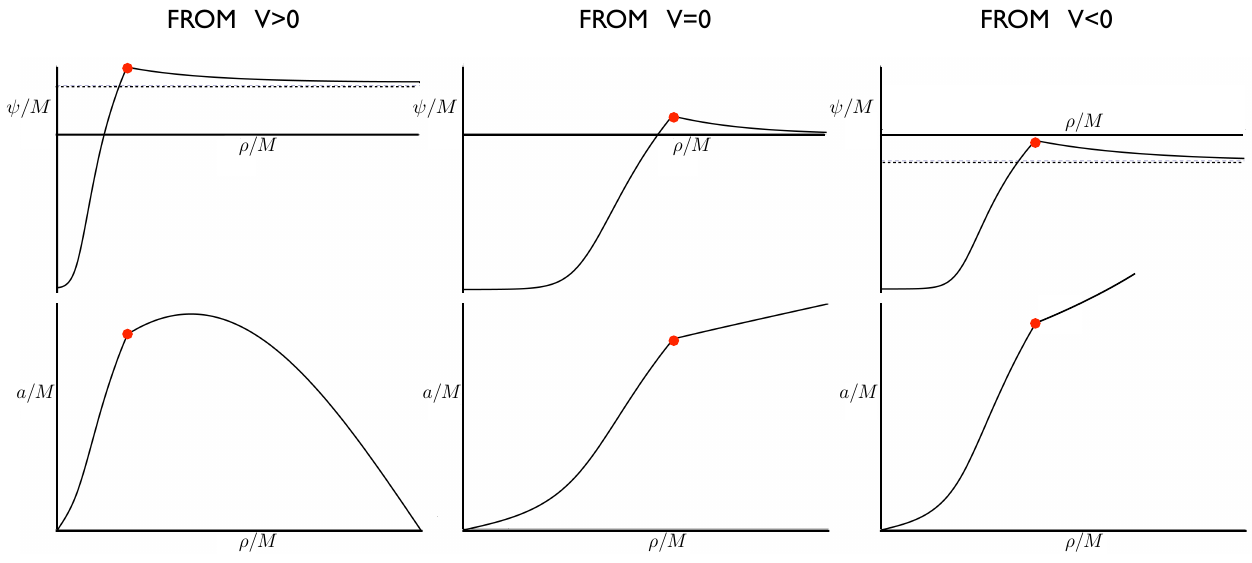} 
    \caption{Sample instanton profiles for flux tunneling in multiflux landscapes, between the same true and false vacua as Fig.~\ref{SampleInstantonMono}.  The corresponding brane tension is larger than in the monoflux case, which has two consequences. First, the bubble must be bigger to compensate for the increased surface tension, so $\bar{\rho}$ and $a(\bar{\rho})$ are bigger. Second, the radion spikes more in the vicinity of the brane, and to such an extent that $\psi(\rho)$ is nonmonotonic.}
    \label{SampleInstantonMulti}
 \end{figure}

In Fig.~\ref{SampleInstantonMono}, instanton profiles are plotted for three sample monoflux transitions. All transitions are to the same AdS true vacuum, but from different false vacua: respectively, a de Sitter, a Minkowski and an AdS. In all cases, $\psi$ is monotonic in $\rho$; though the effect of the brane tension is to locally spike the radion, this effect is overwhelmed by $\psi$ interpolating between its different vacuum values.

In Fig.~\ref{SampleInstantonMulti}, instanton profiles are now plotted for three sample multiflux transitions. The starting and finishing $V$'s are the same as for the monoflux transitions; the difference is that the brane is heavier in the multiflux case. This has two visible consequences: multiflux bubbles must be bigger to compensate for the bigger surface tension; and the heavier brane induces a spike in the radion that is now so large that $\psi$ is nonmonotonic. (For transitions to deep enough AdS, much deeper than shown here, the change in $\psi$ becomes large enough that it does eventually overwhelm the spike caused by the multiflux brane and the profile becomes monotonic.)

\section{The rate to flux tunnel} \label{sec:Bofn}

The tunneling rate $\Gamma$ between two vacua is given by
\begin{equation}
\Gamma\sim e^{-B/\hbar}, \hspace{.3 in} B=S_E(\text{instanton})-S_E(\text{false vacuum}).
\end{equation}

\begin{figure}[htbp] 
    \centering
    \includegraphics[width=6.5in]{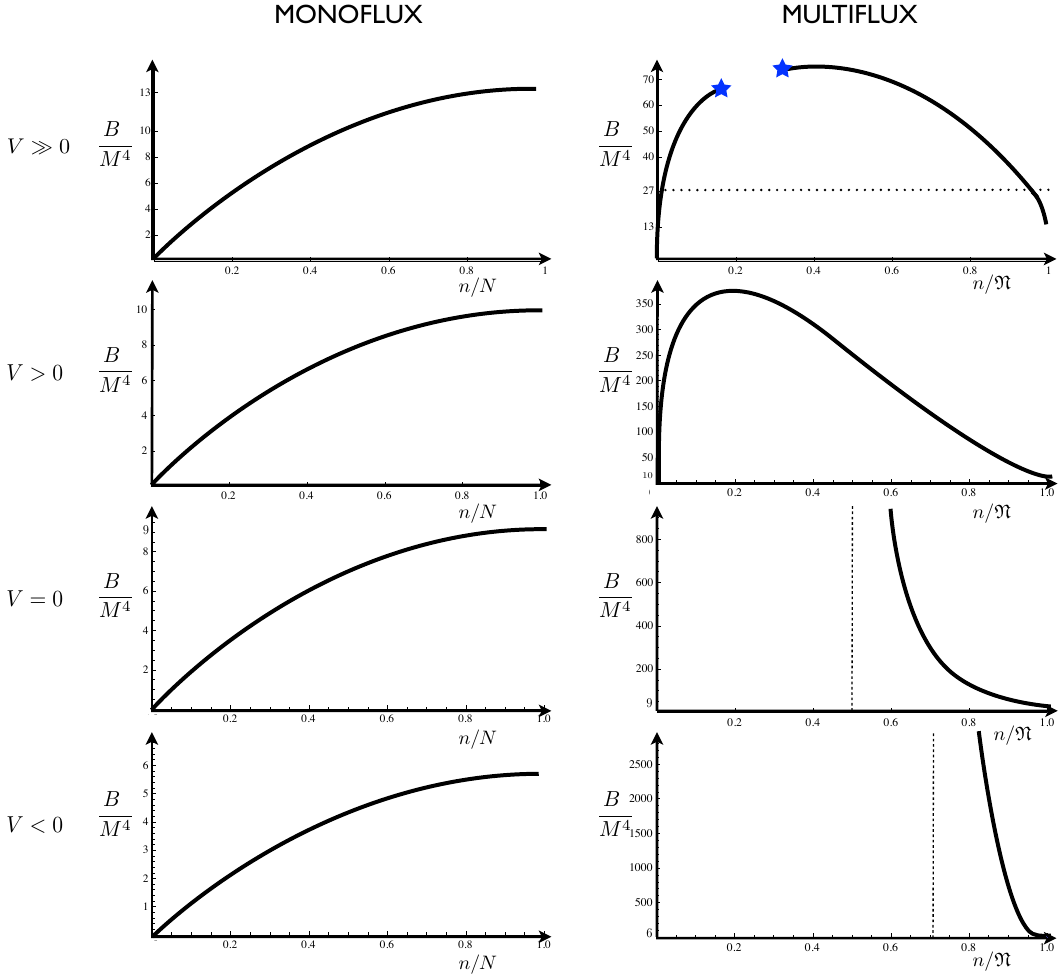} 
    \caption{The tunneling rate $B$ as a function of the number $n$ of flux units dropped. We plot this both for monoflux and multiflux landscapes, and from four different starting points: very high de Sitter ($V = 0.94$), de Sitter ($V=0.28 $), Minkowski ($V=0$) and AdS ($V=-3.55$). Discrete true vacua lie along the curves with spacing set by $g$.  For monoflux decays, the curves rise monotonically with $n$. For multiflux decays from AdS and Minkowski, $B(n)$ falls monotonically---in fact there is a critical $n$, indicated by the vertical dashed line, at which $B(n)$ diverges; decays with smaller $n$ are gravitationally forbidden. For multiflux decays from de Sitter, $B(n)$ rises sharply and then falls off again. For multiflux decays from very high de Sitter, decompactification is competitive and its rate is indicated by the dotted horizontal line; it so dominates flux decay for intermediate $n$ that no instanton exists in the range between the two stars.  Monoflux transitions are faster than the corresponding multiflux transitions, with the exception of the maximum value of $n$; they agree on the rate of the bubble of nothing.  The Minkowski bubble of nothing has tunneling exponent $B\sim9.07M^4$.  Bubbles of nothing from de Sitter have larger $B$ and those from AdS have smaller $B$.}
    \label{Bofn}
 \end{figure}

In Fig.~\ref{Bofn}, the tunneling exponent $B$ is plotted for decays from four false vacua: very high de Sitter, intermediate de Sitter, Minkowski and AdS. It is shown as a function of the fraction of the landscape jumped: $n/N$ for a monoflux landscape and $n/\mathfrak{N}$ for a multiflux landscape.  Though we have drawn it as a solid line, true vacua only lie at discrete points along this curve, with a spacing set by $g$. (Other than this, the values of $g$ and $\Lambda_6$, have no effect.)

For monoflux landscapes, the plots look similar for each starting point. $B$ increases monotonically and then flattens out near $n = N$. Though in the thin-wall approximation giant leaps appear the most likely decay \cite{paper1}, this full calculation shows that actually small \nopagebreak steps dominate. 

For multiflux landscapes, it depends where you start. From Minkowski and AdS vacua, $B$ is monotonically \emph{decreasing} so that giant leaps are preferred and the best is to drop all units of flux (which leads to a bubble of nothing, as we will discuss in more detail in the next section). Small enough steps aren't just suppressed, they're forbidden: a whole section of the landscape is inaccessible. An interpretation of this behavior was given by Coleman and De Luccia \cite{CDL}. Quantum decay conserves energy, so the bubble's surface area to volume ratio must be small enough that the energy benefit of the interior can compensate for the energy cost of the surface tension. But because the interior is negatively curved, its volume scales asymptotically with its area, and the surface area to volume ratio cannot be made arbitrarily low.  For small steps, the bubble can never be big enough to have zero energy and the instantons are gravitationally blocked.

From de Sitter vacua, no decay is gravitationally blocked. $B$ rises fast at small $n$ and then turns over and falls slowly down towards $n = \mathfrak{N}$. Much the same is true from very high de Sitter, but with a complication. From very high de Sitter decompactification so dominates flux tunneling for intermediate $n$ that the flux tunneling instanton disappears in the range between the stars. This is not gravitational blocking---the instanton has instead been swallowed by a much faster decay.  Similar disappearing instantons have been discussed elsewhere in the literature \cite{paper1, DisappearingInstanton}; we  give a general account of instanton disappearances and a complete account of this particular disappearance in \cite{TheCaseIsCracked}.  From even higher de Sitter, the stars move apart, eventually swallowing even the bubble of nothing. 

If you start very near the top of the multiflux landscape, the most likely decay is decompactification. From slightly farther down the landscape, it depends on $g$, which controls the proximity of the next vacuum. For a given de Sitter space, there is a critical value of $g$ for which the rate to nucleate a single brane is the same as the rate to nucleate a bubble of nothing. At smaller $g$, the first vacuum appears earlier and has smaller $B$, so that the optimal decay path is a small step. At larger $g$, giant leaps dominate. This critical value of $g$ decreases as you move down the landscape, so that for fixed $g$, if you start near the top you expect to first take small steps and then, still well within the de Sitter regime, make a giant leap to a bubble of nothing. 

Two approximations go into these plots. The brane breaks the spherical symmetry of the extra dimensions. Our first approximation is to ignore the effects of this breaking---we treat the shape moduli as fixed and ignore higher Kaluza-Klein modes of the flux. Though we have eliminated the thin-wall approximation of \cite{paper1}, our second approximation is to continue to treat the brane as thin. This allows us to use the membrane action Eq.~\ref{eq:brane}.   
This approximation is reliable at small and intermediate $n$, establishing that multiflux giant leaps beat small steps; but it gets steadily worse as we tunnel farther and by the bubble of nothing solution the thickness of the brane ($r_0 = 3T/16 \pi$)  is approaching the size of the extra dimensions and the proper size of the bubble.  We anticipate that moving away from this approximation will increase $B$ for the largest values of $n$.  We could move away from this approximation by committing to a particular model that resolves the core of the brane, as in \cite{fluxnothing}; such a model would make the brane tension, and also its thickness, into tunable parameters.  

Monoflux transitions are faster than the equivalent multiflux transitions, because for a given $\Delta F^2$, the brane tension is smaller. But as $\Delta F^2$ increases, monoflux transitions get slower and multiflux transitions generally get faster. In the limit that all the flux is dropped, the rates are the same: the transitions have the same $\Delta F^2$, the same brane tension, and the same instanton profile. This limit corresponds to the bubble of nothing.

\section{Bubble of nothing} \label{sec:BubbleOfNothing}

The fastest decay in a multiflux landscape is often achieved in the limit $n \rightarrow \mathfrak{N}$, so that all the flux is discharged, the radion is no longer stabilized against collapse, both $\psi_\text{min}$ and $V_\text{min} \rightarrow - \infty$, the extra dimensions shrink to zero size, spacetime pinches off, and we create a bubble of nothing. 

\subsection{Comparison to unstabilized 5D bubble of nothing}

\begin{figure}[h!] 
    \centering
    \includegraphics[width=\textwidth]{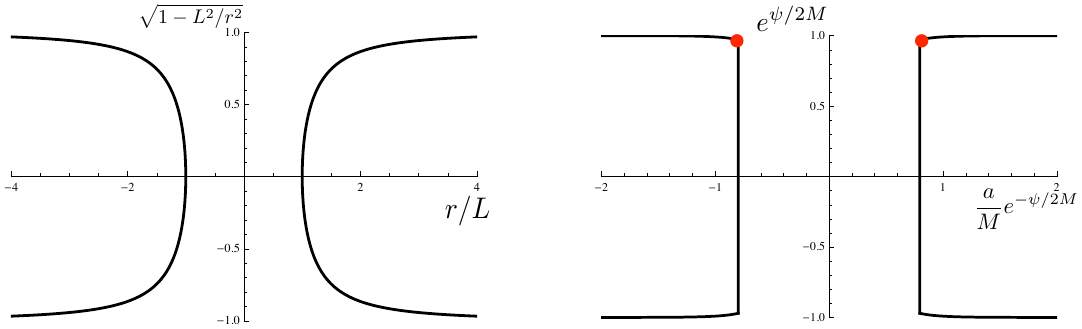} 
    \caption{Left pane: a cross-section through the unstabilized 5D bubble of nothing, plotting the normalized proper size of the extra dimension against the normalized proper area-radius. Right pane: a similar cross-section through the stabilized 6D bubble of nothing. The brane is indicated by a red dot. Inside the brane, the extra dimensions shrink so rapidly to zero that the profile appears almost vertical.}
    \label{fig:cross}
 \end{figure}

For comparison, we review the original bubble of nothing, discovered by Witten \cite{Witten}, in which an unstabilized extra dimension shrinks to zero. The instanton has metric
\begin{equation}
ds^2 = \frac{dr^2}{1 - \frac{L^2}{r^2}} + r^2 d\Omega_3^{\,2} + \left( 1 - \frac{L^2}{r^2} \right) L^2 d\varphi^2, \label{eq:Wittencoordinates}
\end{equation}
 with $L<r<\infty$ and $0< \varphi<2 \pi$. 
 The left pane of Fig.~\ref{fig:cross} shows a cross-section through this bubble at the instant of nucleation. The size of the extra dimension, $\sqrt{1 - L^2 / r^2}$, is plotted against the area-radius, $r$. The extra dimension shrinks smoothly to zero at $r=L$, and spacetime pinches off: for $r<L$ there is literally nothing. After nucleation, the bubble accelerates outwards approaching the speed of light, and the hole in spacetime grows.

This bubble of nothing describes the collapse of a \emph {single}, \emph{unstabilized} extra dimension in Minkowski. Bubbles of nothing in 6D Einstein-Maxwell theory describe the collapse of \emph{two}, \emph{stabilized} extra dimensions from any of the Minkowski, de Sitter or AdS vacua. 
The right pane of Fig.~\ref{fig:cross} shows a cross-section through the bubble of nothing from the Minkowski vacuum. The metric is
\begin{equation}
ds^2=e^{-\psi/M} \left( d \rho^2 + a^2 d \Omega_3^{\,2} \right)  + e^{\psi/M} R^2  d\Omega_2^{\,2},
\label{eq:6Dmetric2}
\end{equation}
so the proper size of the extra dimensions is $e^{\psi / 2M}$ and the proper area-radius is $\frac{a}{M} e^{- \psi / 2M}$. The brane sits at finite $\psi < \psi_\text{false}$; inside the brane, there is no stabilizing flux and the size of the extra dimensions shrinks to zero. It does so much more precipitously than in the 5D case, as there is now a curvature term driving the collapse. Defining the pinch-off point to be $\rho=0$, we can expand the solutions to Eqs.~\ref{eompsi} and \ref{eomgrav} at small $\rho $ to give
\begin{eqnarray}
e^{- \psi/M} & = & \frac{1}{ 4 \pi^{\frac{1}{2}}} \frac{M}{\rho} + \cdots \\
\frac{ a}{M}  & = & \beta \left( \frac{ \rho }{M} \right)^{\frac{1}{2}} + \cdots \label{eq:arho}
 \end{eqnarray}
Any value of $\beta$ solves the equations of motion, but only one value ensures that $\psi$ settles in at the desired value at infinity, rather than under-shooting or over-shooting. We can find this value numerically and for decays from Minkowski it is $\beta \sim 2.12307$.\footnote{The bubble of nothing solution given here differs from that given in \cite{fluxnothing}.  There, a Higgs potential is used to resolve the core of the brane, and an ansatz is assumed that puts the center of the core at $\psi\rightarrow-\infty$.
Core-resolution is a desirable (albeit model-dependent) way to move beyond our thin-brane approximation.  But the substantive difference is that our thin brane sits at finite $\psi$, which is inconsistent with the ansatz, suggesting the solution in \cite{fluxnothing} has too many negative modes---we anticipate an instability whereby the points near $\rho=0$ roll off the crest of the Higgs potential and the brane shifts to larger $\psi$.  Combining a core-resolution technique with a more general ansatz would be definitive, though computationally intensive.}  

The proper area of the $\rho = 0$ equatorial two-sphere is $4\pi a^2e^{-\psi/M}=\pi^{1/2} \beta^2 M^2$, which is finite even though $a \rightarrow 0$ and $\psi \rightarrow - \infty$. The solution is smooth despite not having $\psi'(0)=0$, as can be seen by defining $x= M/\sqrt{4\pi}\times e^{\psi / 2M}$ and expanding near $x=0$ to give
\begin{equation}
ds^2 = dx^2 + \frac{\pi^{\frac{1}{2}} \beta^2 M^2}{4 \pi } d\Omega_3^{\,2} + x^2 d\Omega_2^{\,2} .
\end{equation}

\subsection{Bubble of nothing as the limit of flux tunneling} \label{subsec:us}
The bubble of nothing in 6D Einstein-Maxwell theory is not an isolated solution---it can be smoothly approached as the limit of flux tunneling in which all the flux is removed. Studying this limit gives new insight into the bubble's structure. 

    \begin{figure}[htp] 
    \centering
    \includegraphics[width=3.5in]{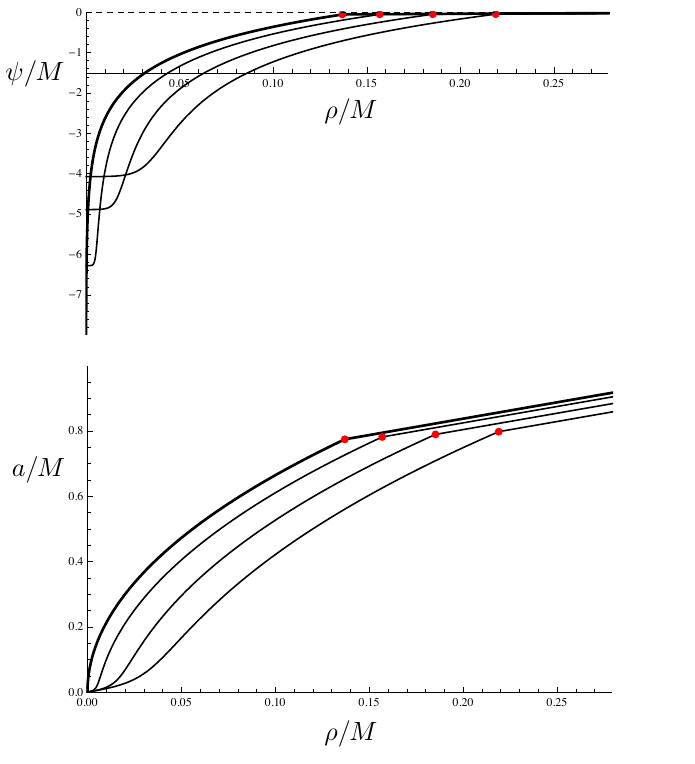} 
        \includegraphics[width=3.8in]{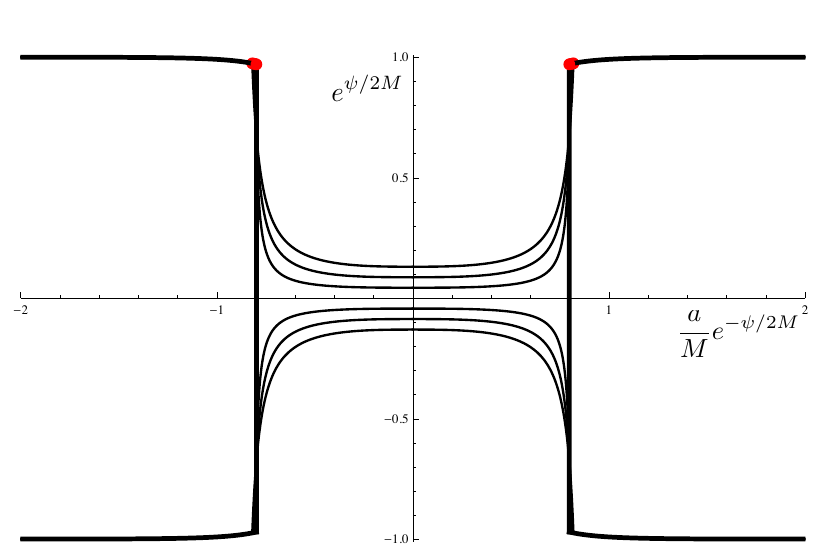} 
    \caption{A sequence of instantons that mediate decay from Minkowski space, and describe flux tunneling by an amount $n = 0.85N$, $n=0.9N$, $n=0.95N$, and, in bold, the bubble of nothing, $n=N$. The plot is made for a monoflux landscape; a multiflux landscape approaches the same limit from a slightly different direction.  On the outside of the bubble, $\psi \rightarrow 0$ so $e^{\psi/2M} \rightarrow 1$. On the inside of the bubble, the extra dimensions relax to their true vacuum size.}  
    \label{fig:bubofnot3}
 \end{figure}

Figure~\ref{fig:bubofnot3}  shows a sequence of solutions that tunnel ever farther down the landscape and approach the bubble of nothing. The instantons can be roughly divided into three segments. In the inner segment, when $\rho$ is small, $\psi$ sits close to its true vacuum value and $a \sim \sinh (\rho / \ell_\text{curv})$, where $\ell_\text{curv}$ is the appropriate AdS curvature length. In the middle segment, $\psi$ interpolates from close to the true vacuum to close to the false vacuum and $a \sim \rho^{1/2}$, as in Eq.~\ref{eq:arho}. In the outer segment, on the other side of the brane, $\psi$ slowly relaxes to its false vacuum value as $\rho \rightarrow \infty$ and, because the plots are for a Minkowski false vacuum, $a \sim \rho$. 
 
As the instantons approach the bubble of nothing, the inner segment gets deeper but shorter---the solution spends less long loitering in the vicinity of the true vacuum (as measured by $\rho$). In the limit that we tunnel all the way down, $\psi_\text{min}$ disappears off to minus infinity, the inner segment vanishes entirely, and the middle segment stretches all the way to $\rho=0$. 

Because $\psi\rightarrow-\infty$, the interior of the bubble has no volume in the \emph{extra} dimensions.  But it's more than that: it also has no volume in the \emph{extended} directions (it's not like a flattened pancake with 3-volume but no thickness).  The interior of the bubble, deep in AdS, is negatively curved, so that $a\sim\sinh(\rho/\ell_\text{curv})$; because this is exponentially growing, most of the volume sits within $\ell_\text{curv}$ of the surface.  In the limit, $V \rightarrow - \infty$ so $\ell_\text{curv}\rightarrow0$ and there is surface area (it's a ``bubble'') but no volume (``of nothing").

\section{Discussion}
\label{sec:Conclusions}
As discussed here and in \cite{paper1}, there are two radion-related effects that help giant leaps in both monoflux and multiflux landscapes.
\begin{enumerate}
\item {\bf Tension effect.} In Einstein frame, the 4D effective brane tension shrinks as the radion swells, but at a cost: to swell, the radion must move away from its minimum.  A second brane at the same point accrues the same benefit, without having to pay again the cost swelling the extra dimensions. Thus the radion mediates an attractive force between branes, reducing the tension of large stacks.
\item $\mathbf {\Delta V}$ {\bf  effect.} For a given $\psi$, $V(\psi)$ changes linearly with $F^2$, but as $F^2$ changes so too does the $\psi$-location of the minimum. $V_\text{min}$ is thus a concave-down function of $F^2$, so that $\Delta V$ grows faster than $\Delta F^2$ (indeed $V_\text{min} \rightarrow - \infty$ as $F^2 \rightarrow 0$). 
\end{enumerate}
There are three effects that make giant leaps more favored in multiflux landscapes than in monoflux landscapes. 
\begin{enumerate}
\item  {\bf Tension effect.}  For monoflux branes, the magnetic repulsion cancels the gravitational attraction, and the tension of a stack grows like $n$ (ignoring the radion).  For multiflux branes, charged under different fluxes, there is no magnetic repulsion, and the tension grows like $\sqrt{n}$.
\item  $\mathbf{\Delta V}$ {\bf  effect.} For monoflux decays, at each step you shed the same $F$, but ever less $F^2$: $\Delta F^2/g^2 = N^2 - (N - n)^2$$=n(2N-n)$. But for multiflux decays, the flux lines don't interact, and at each step you shed the same $F^2$: $\Delta F^2/g^2 = \mathfrak{N} - (\mathfrak{N}-n) = n$.
\item {\bf Thick-wall effect.}  Giant leaps have thick walls, and thick walls slow decay---they encroach upon the interior true vacuum and increase $B$. For monoflux transitions, which proceed by small bubbles, this effect is large and in the end sinks giant leaps. However, as we saw in Fig.~\ref{SampleInstantonMulti}, multiflux transitions proceed by bigger bubbles, which can accommodate the thick walls with a proportionately smaller increase in $B$. 
\end{enumerate}
Incorporating all these effects, we have calculated the tunneling rate as a function of $n$, in the thin-brane approximation.  This allows us to track the most likely decay chain. From the highest de Sitter states, the fastest decay is to decompactify. From somewhat lower down, it depends on the type of landscape. For monoflux landscapes, built of many units of a single flux, the most likely decay chain is to discharge one unit at a time, proceeding by small steps all the way to the bottom. For multiflux landscapes, built of a single unit each of many fluxes, high de Sitter states take a few small steps before a catastrophic giant leap all the way to a bubble of nothing.

What about intermediate cases, with many units of many fluxes? Different types of flux still like to discharge together, because they enjoy all the benefits described above. That said, if the $N_i$ all share a common factor $p$, then the bubble of nothing ($n_i=N_i$) is necessarily subdominant.  Because monoflux landscapes take small steps, we know that tunneling by $n_i=N_i/p$ must be faster.  If the $N_i$ don't share a common factor and the $g_i$ vary, then the cases proliferate, but for much of the landscape, giant leaps will still be the fastest decay.

If, as seems plausible, we ourselves live in a multiflux landscape, then we draw two conclusions. We arrived here by an exponentially subdominant decay, and we will leave here by a bubble of nothing.

\section*{Acknowledgements}
Thanks to Paul Steinhardt and Herman Verlinde for useful feedback and to Jose Juan Blanco-Pillado, Handhika Ramadhan and Benjamin Shlaer for helpful discussion on the bubble of nothing.  Hat tip to Kendrick and Mari, much respect.
\appendix 
\section*{Appendix: a new instanton} \label{sec:Appendix} 

In Sec.~4, we considered both flux tunneling instantons and decompactification instantons.  But there is one more possibility. In this Appendix, we construct the instanton that {simultaneously} discharges flux \emph{and} decompactifies.  

\begin{figure}[htbp] 
   \centering
   \includegraphics[width=6in]{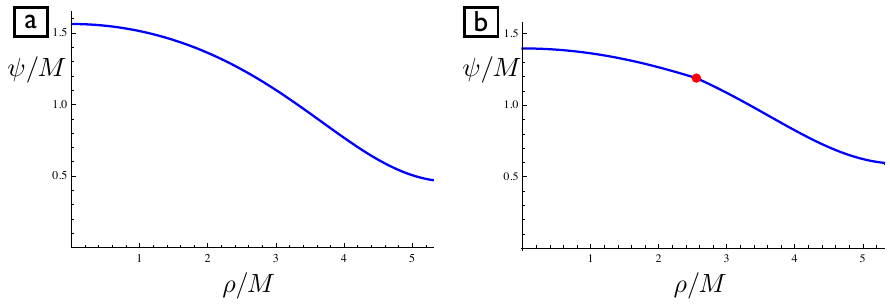} 
   \caption{(a) The instanton that mediates decompactification out of a de Sitter vacuum. (b) The instanton that mediates simultaneous decompactification and flux tunneling out of the same de Sitter vacuum. The brane discharges flux on the inside, so that the field is still unstable to decompactification even though $\psi(0)$ is not as large. Similar instantons exist that discharge more or less flux. }
   \label{fig:decompactification}
\end{figure}

In Fig.~\ref{fig:decompactification}a we plot the standard decompactification instanton. This instanton doesn't change the flux, and instead nucleates a bubble inside of which $\psi$ is over the hump in the effective potential of Fig.~1 and will classically roll $\psi \rightarrow \infty$.
In Fig.~\ref{fig:decompactification}b we plot the new instanton. As before, $\psi$ is over the hump, so this instanton corresponds to decompactification. But there is now a charged brane, so that there is less flux on the inside, the effective potential is reduced, and hump is shrunk. 

\begin{figure}[htbp] 
   \centering
   \includegraphics[width=4in]{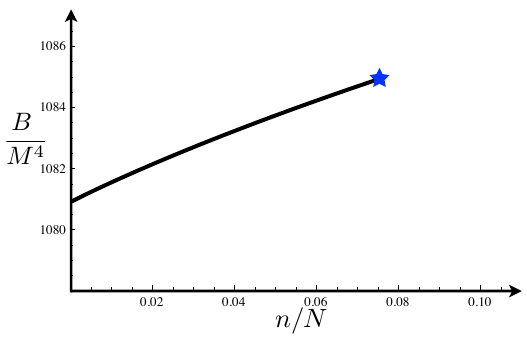} 
   \caption{The rate to simultaneously discharge flux \emph{and} decompactify, as a function of the proportion of flux dropped $n / N$. At $n=0$, no flux is dropped, so this is just standard decompactification. Since the rate grows with $n$, simultaneously dropping flux and decompactifying is subdominant to standard decompactification. Similar plots, with identical conclusions, can be drawn from other de Sitter false vacua and for multiflux landscapes.}
   \label{fig:Bdecompactification}
\end{figure}

Which method of decompactifying is fastest? On the one hand the new method lowers the hump in the effective potential, making decompactification easier. On the other hand, it requires the nucleation of a brane, making decompactification harder. Figure~\ref{fig:Bdecompactification} shows that this second effect is strongest, so that the new instantons are always subdominant to standard decompactification.   This figure shows that decompactification gets slower the more flux is simultaneously dropped. 

Indeed, standard decompactification so dominates this new instanton, that eventually the new instanton disappears at a critical $n < N$ (just as the regular flux tunneling instanton disappears from a high enough de Sitter state, as in the top left pane of Fig.~4).   See \cite{TheCaseIsCracked} for an account of the disappearance.  As the false vacuum is raised, the star moves left and eventually swallows everything, including even the standard decompactification instanton. (At the moment the standard decompactification instanton is swallowed, it is replaced by a Hawking-Moss instanton\cite{HawkingMoss, Jensen:1988zx}.)


\begin{thebibliography}{99}

\bibitem{paper1}
 A.~R.~Brown and A.~Dahlen,
 ``Small Steps and Giant Leaps in the Landscape,''
 Phys.\ Rev.\ D {\bf 82}, 083519 (2010)
  [arXiv:1004.3994 [hep-th]].

\bibitem{paper2}
  A.~R.~Brown and A.~Dahlen,
  ``Giant Leaps and Minimal Branes in Multi-Dimensional Flux Landscapes,''
  Phys.\ Rev.\  D {\bf 84}, 023513 (2011)
  [arXiv:1010.5241 [hep-th]].

 \bibitem{Witten}
  E.~Witten,
  ``Instability Of The Kaluza-Klein Vacuum,''
  Nucl.\ Phys.\  B {\bf 195}, 481 (1982).
  

\bibitem{ChangeD}
  A.~D.~Linde and M.~I.~Zelnikov,
  ``Inflationary Universe with Fluctuating Dimension,''
    Phys.\ Lett.\  B {\bf 215}, 59 (1988);

  S.~M.~Carroll, M.~C.~Johnson and L.~Randall,
  ``Dynamical compactification from de Sitter space,''
  JHEP {\bf 0911}, 094 (2009)
  [arXiv:0904.3115 [hep-th]];
  
    J.~J.~Blanco-Pillado, D.~Schwartz-Perlov and A.~Vilenkin,
  ``Transdimensional Tunneling in the Multiverse,''
  JCAP {\bf 1005}, 005 (2010)
  [arXiv:0912.4082 [hep-th]].

\bibitem{Feng:2000if}
  J.~L.~Feng, J.~March-Russell, S.~Sethi and F.~Wilczek,
  ``Saltatory relaxation of the cosmological constant,'
  Nucl.\ Phys.\  B {\bf 602}, 307 (2001)
  [arXiv:hep-th/0005276].

\bibitem{perc1}
A.~H.~Guth and E.~J.~Weinberg,
  ``Could The Universe Have Recovered From A Slow First Order Phase
  Transition?,''
  Nucl.\ Phys.\  B {\bf 212}, 321 (1983).

\bibitem{perc2}
  S.~Sarangi, G.~Shiu and B.~Shlaer,
  ``Rapid Tunneling and Percolation in the Landscape,''
  Int.\ J.\ Mod.\ Phys.\  A {\bf 24}, 741 (2009)
  [arXiv:0708.4375 [hep-th]].
 
\bibitem{FreundRubin} 
  P.~G.~O.~Freund and M.~A.~Rubin,
  ``Dynamics Of Dimensional Reduction,''
  Phys.\ Lett.\  B {\bf 97}, 233 (1980).

\bibitem{BSV}
  J.~J.~Blanco-Pillado, D.~Schwartz-Perlov and A.~Vilenkin,
  ``Quantum Tunneling in Flux Compactifications,''
  JCAP {\bf 0912}, 006 (2009)
  [arXiv:0904.3106 [hep-th]].
  
 \bibitem{Yang}
  I.~S.~Yang,
   ``Stretched extra dimensions and bubbles of nothing in a toy model
  landscape,''
  Phys.\ Rev.\  D {\bf 81}, 125020 (2010)
  [arXiv:0910.1397 [hep-th]].
  
  \bibitem{CDL}
 S.~R.~Coleman and F.~De Luccia,
  ``Gravitational Effects On And Of Vacuum Decay,''
  Phys.\ Rev.\  D {\bf 21}, 3305 (1980).

\bibitem{DisappearingInstanton}
  M.~Cvetic and H.~H.~Soleng,
  ``Naked singularities in dilatonic domain wall space times,''
  Phys.\ Rev.\  D {\bf 51}, 5768 (1995)
  [arXiv:hep-th/9411170];

  M.~C.~Johnson and M.~Larfors,
  ``An obstacle to populating the string theory landscape,''
  Phys.\ Rev.\  D {\bf 78}, 123513 (2008)
  [arXiv:0809.2604 [hep-th]];
  
  A.~Aguirre, M.~C.~Johnson and M.~Larfors,
  ``Runaway dilatonic domain walls,''
  Phys.\ Rev.\  D {\bf 81}, 043527 (2010)
  [arXiv:0911.4342 [hep-th]].
  
  \bibitem{TheCaseIsCracked}
  A.~R.~Brown and A.~Dahlen,
  ``The Case of the Disappearing Instanton,''
  arXiv:1106.0527 [hep-th].
  
\bibitem{fluxnothing}
  J.~J.~Blanco-Pillado and B.~Shlaer,
  ``Bubbles of Nothing in Flux Compactifications,''
Phys.\ Rev.\  D {\bf 82}, 086015 (2010)
  [arXiv:1002.4408 [hep-th]].
    
  J.~J.~Blanco-Pillado, H.~S.~Ramadhan and B.~Shlaer,
  ``Decay of flux vacua to nothing,''
 JCAP {\bf 1010}, 029 (2010)
  [arXiv:1009.0753 [hep-th]].
  
\bibitem{HawkingMoss}
  S.~W.~Hawking and I.~G.~Moss,
  ``Supercooled Phase Transitions In The Very Early Universe,''
  Phys.\ Lett.\  B {\bf 110}, 35 (1982).

\bibitem{Jensen:1988zx}
  L.~G.~Jensen and P.~J.~Steinhardt,
  ``Bubble Nucleation for Flat Potential Barriers,''
  Nucl.\ Phys.\  B {\bf 317}, 693 (1989).
  
  
\end{thebibliography}
\end{document}